\documentclass[11pt,a4paper,twoside,groupcitations]{article}
\usepackage[T1]{fontenc}
\usepackage[ansinew]{inputenc}
\usepackage[english]{babel}
\usepackage{amsfonts}
\usepackage{amsmath}
\usepackage{array}
\usepackage{amsthm}
\usepackage{amssymb}
\usepackage{graphicx}
\usepackage{subfigure}
\usepackage{braket}
\usepackage{eucal}
\usepackage{verbatim}
\usepackage[table]{xcolor}
\usepackage{caption}
\usepackage{multicol}
\usepackage{tikz}
\usetikzlibrary{positioning,arrows}
\usetikzlibrary{decorations.pathmorphing}
\usetikzlibrary{decorations.markings}
\usetikzlibrary{calc,decorations.markings}
\usetikzlibrary{arrows,shapes}
\usetikzlibrary{matrix,arrows}
\usepackage{pgfplots}
\usepackage{xparse}
\definecolor{jade}{HTML}{00A86B}
\newcommand{\be}{\begin{eqnarray}}
\newcommand{\ee}{\end{eqnarray}}
\newcommand{\pro}[2]{\mbox{$\langle\, #1 \mid #2\,\rangle$}}
\newcommand{\expec}[1]{\mbox{$\langle\, #1\,\rangle$}}

\renewcommand{\a}{\hat a}
\newcommand{\ac}{\hat a^{\dagger}}

\renewcommand{\d}{\mbox{${\rm d}$}} 
\newcommand{\lp}{\ell_{\rm p}}
\newcommand{\mpl}{m_{\rm p}}
\newcommand{\gn}{G_{\rm N}}

\newcommand{\Rh}{R_{\rm H}}

\newcommand{\qb}{q_{\rm B}}
\newcommand{\qf}{q_\Phi}
\newcommand{\jb}{J_{\rm B}}

\newcommand{\erf}{{\rm erf}}
\title{\bf Quantum corpuscular corrections to the Newtonian potential}
\author{Roberto~Casadio$^{ab}$\thanks{E-mail: casadio@bo.infn.it}
$\ ,$
Andrea~Giugno$^{c}$\thanks{E-mail: A.Giugno@physik.uni-muenchen.de}
$\ ,$
Andrea~Giusti$^{ab}$\thanks{E-mail: andrea.giusti@bo.infn.it}
$\ $
and
Michele~Lenzi$^{a}$\thanks{E-mail: michele.lenzi@studio.unibo.it}
\\
\\
$^a${\em Dipartimento di Fisica e Astronomia, Universit\`a di Bologna}
\\
{\em via Irnerio~46, 40126 Bologna, Italy}
\\
\\
$^b${\em I.N.F.N., Sezione di Bologna, I.S.~FLAG}
\\
{\em viale B.~Pichat~6/2, 40127 Bologna, Italy}
\\
\\
$^c${\em Arnold Sommerfeld Center, Ludwig-Maximilians-Universit\"at}
\\
{\em Theresienstra{\ss}e~37, 80333 M\"unchen, Germany}
}
\begin{document}
\maketitle
\begin{abstract}
We study an effective quantum description of the static gravitational potential for
spherically symmetric systems up to the first post-Newtonian order.
We start by obtaining a Lagrangian for the gravitational potential coupled to a static
matter source from the weak field expansion of the Einstein-Hilbert action.
By analysing a few classical solutions of the resulting field equation,
we show that our construction leads to the expected post-Newtonian expressions.
Next, we show that one can reproduce the classical Newtonian results very
accurately by employing a coherent quantum state and modifications to include
the first post-Newtonian corrections are considered. 
Our findings establish a connection between the corpuscular model of black holes
and post-Newtonian gravity, and set the stage for further investigations of these
quantum models.
\par
\null
\par
\textit{PACS - 04.70.Dy, 04.70.-s, 04.60.-m}
\end{abstract}
\section{Introduction}
\setcounter{equation}{0}
\label{Sintro}
In the Newtonian theory, energy is a well-defined quantity and is conserved along physical
trajectories (barring friction), which ensures the existence of a scalar potential for the gravitational force.
In General Relativity~\cite{weinberg}, the very concept of energy becomes much more problematic
(see, e.g.~\cite{faraoni} and References therein) and there is no invariant notion of a scalar potential.
Even if one just considers the motion of test particles, the existence of conserved quantities along
geodesics requires the presence of Killing vector fields. 
In sufficiently symmetric space-times, one may therefore end up with equations of motion containing
potential terms, whose explicit form will still depend on the choice of observer (time and spatial coordinates).
Overall, such premises allow for a ``Newtonian-like'' description of gravitating systems with strong
space-time symmetries, like time-independence and isotropy, which can in turn be quantised by standard
methods~\cite{duff,donoghue}.
\par
We are here particularly interested in static and isotropic compact sources, for which one can indeed
determine an effective theory for the gravitational potential, up to a certain degree of confidence.
When the local curvature of space-time is weak and test particles propagate
at non-relativistic speed, non-linearities are suppressed.
The geodesic equation of motion thereby takes the form of the standard Newtonian law with a potential
determined by the Poisson equation, and Post-Newtonian corrections can be further obtained by including
non-linear interaction terms.
The inclusion of these non-linear terms in the quantum effective description of the gravitational potential 
are precisely what we are going to address in this work, following on the results of Ref.~\cite{Casadio:2016zpl}.
\par
One of the motivations for this study is provided by the corpuscular model of gravity recently theorised by
Dvali and Gomez~\cite{DvaliGomez}.
According to this model, a black hole is described by a large number of gravitons in the same (macroscopically
large) state,
thus realising a Bose-Einstein condensate at the critical point~\cite{flassig,becBH}. 
In particular, the constituents of such a self-gravitating object are assumed to be marginally bound
in their gravitational potential well~\footnote{For improvements on this approximation,
see Refs.~\cite{qhbh,mueck,kerr}.}, whose size is given by the characteristic Compton-de~Broglie
wavelength $\lambda_{\rm G}\sim \Rh$, where~\footnote{We shall mostly use units with $c=1$ and the Newton
constant $\gn=\lp/\mpl$, where $\lp$ is the Planck length and $\mpl$ the Planck mass (so that $\hbar=\lp\,\mpl$).}
\be
\Rh=2\,\lp\,\frac{M}{\mpl}
\label{Rh}
\ee
is the Schwarzschild radius of the black hole of mass $M$, and whose depth is proportional to
the very large number $N_{\rm G}\sim M^2/\mpl^2$ of soft quanta in this condensate~\cite{ruffini}.
In the original proposal~\cite{DvaliGomez}, the role of matter was argued to be essentially negligible by considering
the number of its degrees of freedom is subdominant with respect to the gravitational ones,
especially when representing black holes of astrophysical size (see also Ref.~\cite{dvali,kuhnel}).
\par
When the contribution of gravitons is properly related to the necessary presence of ordinary
baryonic matter, not only the picture enriches, but it also becomes clearly connected to
the post-Newtonian approximation~\cite{Casadio:2016zpl}.
The basic idea is very easy to explain:
suppose we consider $N$ baryons of rest mass $\mu$ very far apart,
so that their total ADM energy~\cite{adm} is simply given by $M=N\,\mu$.
As these baryons fall towards each other, while staying inside a sphere of radius $R$,
their (negative) gravitational energy is given by
\be
U_{\rm BG}
\sim
N\,\mu\, V_{\rm N}
\sim
-\frac{\lp\,M^2}{\mpl\,R}
\ ,
\ee
where $V_{\rm N}\sim -{\lp\,M}/{\mpl\,R}$ is the (negative) Newtonian potential.
In terms of quantum physics, this gravitational potential can be represented by the expectation value
of a scalar field $\hat \Phi$ over a coherent state $\ket{g}$,
\be
\expec{g|\,\hat \Phi\,|g}
\sim
V_{\rm N}
\ .
\ee 
The immediate aftermath is that the graviton number $N_{\rm G}$ generated by matter inside
the sphere of radius $R$ is determined by the normalisation of the coherent state and reproduces
Bekenstein's area law~\cite{bekenstein}, that is
\be
N_{\rm G}
\sim 
\frac{M^2}{\mpl^2}
\sim
\frac{\Rh^2}{\lp^2}
\ ,
\ee
where $\Rh$ is now the gravitational radius of the sphere of baryons.
In addition to that, assuming most gravitons have the same wave-length $\lambda_{\rm G}$,
the (negative) energy of each single graviton is correspondingly given by
\be
\epsilon_{\rm G}
\sim
\frac{U_{\rm BG}}{N_{\rm G}}
\sim
-\frac{\mpl\,\lp}{R} 
\ ,
\ee
which yields the typical Compton-de~Broglie length $\lambda_{\rm G}\sim R$.
The graviton self-interaction energy hence reproduces the (positive) post-Newtonian energy,
\be
U_{\rm GG}(R)
\sim
N_{\rm G} \, \epsilon_{\rm G} \, \expec{g|\,\hat \Phi\,|g}
\sim
\frac{\lp^2\,M^3}{\mpl^2\,R^2}
\ .
\ee
This view is consistent with the standard lore, since the $U_{\rm GG}\ll \left|U_{\rm BG}\right|$
for a star with size $R\gg\Rh$.
Furthermore, for $R\simeq \Rh$, one has
\be
U(\Rh)
\equiv
U_{\rm BG}(\Rh)+U_{\rm GG}(\Rh)
\simeq
0
\ ,
\label{maxpack}
\ee
which is precisely the ``maximal packing'' condition of Ref.~\cite{DvaliGomez}.
\par
We shall here refine the findings of Ref.~\cite{Casadio:2016zpl}, by first deriving the effective action
for a static and spherically symmetric potential from the Einstein-Hilbert action in the weak field and
non-relativistic approximations.
We shall then show that including higher order terms yields classical results in agreement with the standard
post-Newtonian expansion of the Schwarzschild metric (see Appendix~\ref{Apost-Newton}) and
a quantum picture overall consistent with the one recalled above from Ref.~\cite{Casadio:2016zpl}.
We remark once more this picture is based on identifying the quantum state of the gravitational potential
as a coherent state of (virtual) soft gravitons, which provides a link between the microscopic dynamics of gravity,
understood in terms of interacting quanta, and the macroscopic description of a curved background. 
\par
The paper is organised as follows:
in Section~\ref{SeffAct}, we construct the effective action for the gravitational potential up to the first
post-Newtonian correction and study a few solutions of the corresponding classical field equation;
the analogous quantum picture is then given in Section~\ref{Squantum}, where we analyse in details
the coherent state and estimate its post-Newtonian corrections;
final considerations and possible future developments are summarised in Section~\ref{Sconc}.
\section{Effective scalar theory for post-Newtonian potential}
\label{SeffAct}
\setcounter{equation}{0}
It is well known that a scalar field can be used as the potential for the velocity of a classical
fluid~\cite{madsen88}.
We will show here that it can also be used in order to describe the usual post-Newtonian
correction that appears in the weak field expansion of the Schwarzschild metric.
It is important to recall that this picture implicitly assumes the choice of a specific reference
frame for static observers (for more details, see Appendix~\ref{Apost-Newton}) 
\par
Let us start from the Einstein-Hilbert action with matter~\cite{weinberg}
\be
S
=
S_{\rm EH}
+
S_{\rm M}
=
\int \d^4 x\,\sqrt{-g}
\left(
-\frac{\mpl}{16\,\pi\,\lp}\,\mathcal{R} +\mathcal{L}_{\rm M}
\right)
\ ,
\label{EHaction}
\ee
where $\mathcal{R}$ is the Ricci scalar and $\mathcal{L}_{\rm M}$ is the Lagrangian density
for the baryonic matter that sources the gravitational field. 
In order to recover the post-Newtonian approximation in this framework, we
must assume the local curvature is small, so that the metric can be written
as $g_{\mu\nu}=\eta_{\mu\nu}+h_{\mu\nu}$, where $\eta_{\mu\nu}$ is the flat
Minkowski metric with signature $(-,+,+,+)$ and $|h_{\mu\nu}|\ll 1$.
The Ricci scalar then takes the simple form
\be
\mathcal{R}
=
\Box h -\partial^\mu \partial^\nu h_{\mu\nu} 
+\mathcal{O}(h^2)
\ ,
\ee
where 
\be
\Box=-\partial_t^2+\triangle
\ee
is the d'Alembertian in flat space, the trace
$h=\eta_{\mu \nu}\,h^{\mu\nu}$, and the linearised Einstein field equation is given by
\be
-\Box h_{\mu\nu}
+\eta_{\mu\nu}\,\Box h 
+\partial_\mu\partial^\lambda h_{\lambda\nu}
+\partial_\nu\partial^\lambda h_{\lambda\mu}
-\eta_{\mu\nu}\,\partial^\lambda\partial^\rho h_{\lambda\rho}
-\partial_\mu\partial_\nu h
=
16\,\pi\, \frac{\lp}{\mpl} \, T_{\mu\nu}
\ .
\label{EinstField}
\ee
In the de~Donder gauge,
\be
2\,\partial^\mu h_{\mu\nu}=\partial_\nu h
\ ,
\label{ddgg}
\ee
the trace of the field equation yields
\be
\Box h
=
16\,\pi\,\frac{\lp}{\mpl}\,T
\ ,
\ee
where $T=\eta^{\mu\nu}\,T_{\mu\nu}$, and Eq.~\eqref{EinstField} reduces to
\be
-\Box h_{\mu\nu}
=
16\,\pi\, \frac{\lp}{\mpl} 
\left(T_{\mu\nu}-\frac{1}{2}\,\eta_{\mu\nu}\,T\right)
\ .
\label{deDonderField}
\ee
\par
In addition to the weak field limit, we assume that all matter in the system moves with
a characteristic velocity much slower than the speed of light in the (implicitly) chosen
reference frame $x^\mu=(t,{\bf x})$.
The only relevant component of the metric is therefore $h_{00}({\bf x})$, and its time
derivatives are also neglected~\footnote{For static configurations, the gauge
condition~\eqref{ddgg} becomes Eq.~\eqref{ddg}, and is always satisfied.}.
The Ricci scalar reduces to
\be
\mathcal{R} 
\simeq 
\triangle h_{00}({\bf x})
\ ,
\ee
and the stress-energy tensor is accordingly determined solely by the energy density
in this non-relativistic regime,
\be
T_{\mu \nu}
=
\frac{2}{\sqrt{-g}}\,
\frac{\delta S_{\rm M}}
{\delta g^{\mu\nu}}
=
2\,\frac{\delta \mathcal{L}_{\rm M}}{\delta g^{\mu\nu}}
-g_{\mu\nu}\, \mathcal{L}_{\rm M}
\simeq
u_{\mu}\,u_{\nu}\,\rho({\bf x})
\ ,
\label{Tij}
\ee
where $u^\mu=\delta^\mu_0$ is the four-velocity of the static source fluid. 
Note further that the above stress-energy tensor follows from the simple
matter Lagrangian
\be
\mathcal{L}_{\rm M}
\simeq
-\rho({\bf x})
\ ,
\label{Lmp0}
\ee
as one can see from the variation of the baryonic matter density~\cite{harko}
\be
\delta\rho
=
\frac{1}{2}\,\rho\left(g_{\mu\nu}+u_\mu\,u_\nu\right)
\delta g^{\mu\nu}
\ ,
\ee
and the well-known formula
\be
\delta\left(\sqrt{-g}\right)
=
-\frac{1}{2}\,\sqrt{-g}\,g_{\mu\nu}\,\delta g^{\mu\nu}
\ .
\ee
This is indeed the case of interest to us here, since we do not consider explicitly the
matter dynamics but only how (static) matter generates the gravitational field in
the non-relativistic limit, in which the matter pressure is negligible~\cite{madsen88}~\footnote{A
non-negligible matter pressure usually complicates the system significantly and
is left for a separate work.}.
In this approximation, Eq.~\eqref{deDonderField} takes the very simple form
\be
\triangle h_{00}({\bf x})
=
-8\,\pi\, \frac{\lp}{\mpl}\, T_{00}({\bf x})
=
-8\,\pi\, \frac{\lp}{\mpl}\,\rho({\bf x})
\ ,
\label{hPoisson}
\ee
since $T_{00}=\rho$ to leading order. 
Finally, we know the Newtonian potential $V_{\rm N}$ is generated by the density $\rho$
according to the Poisson equation
\be
\triangle V_{\rm N}
=
4\,\pi\, \frac{\lp}{\mpl}\, \rho
\ ,
\label{PoissonVN}
\ee
which lets us identify $h_{00}=-2\,V_{\rm N}$.
\par
It is now straightforward to introduce an effective scalar field theory for the gravitational potential.
First of all, we shall just consider (static) spherically symmetric systems, so that $\rho=\rho(r)$ and
$V_{\rm N}=V_{\rm N}(r)$, correspondingly.
We replace the Einstein-Hilbert action $S_{\rm EH}$ in Eq.~\eqref{EHaction} with the massless
Fierz-Pauli action so that, in the approximation~\eqref{Tij} and~\eqref{Lmp0}, we obtain the total
Lagrangian (see Appendix~\ref{EHNLO})
\be
L[V_{\rm N}]
&\!\!\simeq\!\!&
4\,\pi
\int_0^\infty
r^2 \,\d r
\left(
\frac{\mpl}{32\,\pi\, \lp}\,h_{00}\,\triangle h_{00}
+\frac{h_{00}}{2}\,\rho
\right)
\notag
\\
&\!\!=\!\!&
4\,\pi
\int_0^\infty
r^2 \,\d r
\left(
\frac{\mpl}{8\,\pi\,\lp}\,V_{\rm N}\,\triangle V_{\rm N}
-\rho\,V_{\rm N}
\right)
\notag
\\
&\!\!=\!\!&
-4\,\pi
\int_0^\infty
r^2 \,\d r
\left[
\frac{\mpl}{8\,\pi\,\lp}\left(V_{\rm N}'\right)^2
+\rho\,V_{\rm N}
\right]
\ ,
\label{LagrNewt}
\ee
where we integrated by parts~\footnote{The boundary conditions that ensure vanishing
of boundary terms will be explicitly shown when necessary.} and $f'\equiv \d f/\d r$.
Varying this Lagrangian with respect to $V_{\rm N}$, we obtain Eq.~\eqref{PoissonVN}
straightforwardly~\footnote{Were one to identify the Lagrangian density in Eq.~\eqref{LagrNewt}
with the pressure $p_{\rm N}$ of the gravitational field, it would appear the Newtonian
potential has the equation of state $p_{\rm N}=-\rho_{\rm N}/3$~\cite{madsen88}.}.
\par
In order to go beyond the Newtonian approximation, we need to modify the latter functional
by adding non-linearities.
We start by computing the Hamiltonian,
\be
H[V_{\rm N}]
=
-L[V_{\rm N}]
=
4\,\pi
\int_0^\infty
r^2\,\d r
\left(
-\frac{\mpl}{8\,\pi\,\lp}\,V_{\rm N}\,\triangle V_{\rm N}
+\rho\,V_{\rm N}
\right)
\ ,
\label{NewtHam}
\ee 
as follows from the static approximation.
If we evaluate this expression on-shell by means of Eq.~\eqref{PoissonVN}, we get the Newtonian
potential energy
\be
U_{\rm N}(r)
=
2\,\pi
\int_0^r 
{\bar r}^2\,\d {\bar r}
\,\rho(\bar r)\, V_{\rm N}(\bar r)
\ ,
\label{Un}
\ee
which one can view as given by the interaction of the matter distribution enclosed in a sphere
of radius $r$ with the gravitational field.
Following Ref.~\cite{Casadio:2016zpl}, we could then define a self-gravitational source $J_V$ given
by the gravitational energy $U_{\rm N}$ per unit volume.
We first note that
\be
U_{\rm N}(r)
&\!\!=\!\!&
\frac{\mpl}{2\,\lp}
\int_0^r 
{\bar r}^2 \,\d {\bar r}\,
V_{\rm N}(\bar r)\,\triangle V_{\rm N}(\bar r)
\notag
\\
&\!\!=\!\!&
-\frac{\mpl}{2\,\lp}\,
\int_0^r 
{\bar r}^2 \,\d {\bar r}\,
\left[ V_{\rm N}'(\bar r) \right]^2
\ ,
\ee
where we used Eq.~\eqref{PoissonVN} and then integrated by parts discarding boundary
terms.
The corresponding energy density is then given by
\be
J_V(r)
=
\frac{1}{4\,\pi\, r^2}\,\frac{\d}{\d r} U_{\rm N}(r)
=
-\frac{\mpl}{8\,\pi\,\lp}
\left[ V_{\rm N}'(r) \right]^2
\ .
\label{JV}
\ee
The appearance of the above contribution can in fact be found at the next-to-leading order
(NLO) in the expansion of the theory~\eqref{EHaction}.
As is shown in Appendix~\ref{EHNLO}, the current $J_V$
is in particular proportional to the NLO term~\eqref{Jh00} coming from the
geometric part of the action.
Upon including this new source term, together with its matter
counterpart~\eqref{rhoV2} from the expansion of the matter Lagrangian,
we obtain the total Lagrangian in Eq.~\eqref{LagrEps} for a new field $V$, namely
\be
L[V]
&\!\!=\!\!&
4\,\pi
\int_0^\infty
r^2\,\d r
\left[
\frac{\mpl}{8\,\pi\,\lp}\,V\,\triangle V
-\qb\,\rho\,V
+2\,\qf
 \left( \qb\, V\,\rho
-2\,J_V
\right)V
\right]
\notag
\\
&\!\!=\!\!&
4\,\pi
\int_0^\infty 
r^2\,\d r
\left[
\frac{\mpl}{8\,\pi\,\lp}\,V\,\triangle V
-\qb\,V\,\rho
\left(
1-2\,\qf\, V
\right)
+
\frac{\qf\,\mpl}{2\,\pi\,\lp}\,V
\left( V'\right)^2
\right]
\notag
\\
&\!\!=\!\!&
-4\,\pi
\int_0^\infty 
r^2\,\d r
\left[
\frac{\mpl}{8\,\pi\,\lp}
\left(1-4\,\qf\,V\right)
\left(V'\right)^2
+\qb\,V\,\rho
\left(1-2\,\qf\,V\right)
\right]
\ ,
\label{LagrV}
\ee
where the parameters $\qb$ and $\qf$ keep track of the coupling of $V$ with matter
and the self-coupling, respectively (see again Appendix~\ref{EHNLO} for the details).
It is important to remark that, beyond the linear order, the construction of an effective
theory from the Einstein-Hilbert action~\eqref{EHaction} is plagued by inconsistencies
when coupled to matter.
In order to overcome these issues, the NLO has therefore been constructed from
the Pauli-Fierz action so as not to spoil the Newtonian approximation~\cite{deser}. 
We will show in the following that the post-Newtonian correction~\eqref{Upost}
is indeed properly recovered for $\qb=\qf=1$.
\par
The Euler-Lagrange equation for $V$ is given by
\be
0
&\!\!=\!\!&
\frac{\delta \mathcal{L}}{\delta V}
-\frac{\d}{\d r}
\left( \frac{\delta \mathcal{L}}{\delta V'}\right)
\notag
\\
&\!\!=\!\!&
4\,\pi\,r^2
\left[
-\qb\,\rho
+4\,\qb\,\qf\, \rho\, V
+\frac{\qf\,\mpl}{2\,\pi\,\lp}\left(V'\right)^2
\right]
+\frac{\mpl}{\lp}
\left[r^2\,V'\left(1-4\,\qf\,V\right)\right]'
\ ,
\ee
and, on taking into account that $r^2\,\triangle f(r)=\left(r^2 \, f'\right)'$
for spherically symmetric functions, we obtain the field equation
\be
\left( 1-4\,\qf\, V \right)
\triangle V
=
4\,\pi\,\qb\,\frac{\lp}{\mpl}\,\rho
\left(1-4\,\qf\,V\right)
+2\,\qf \left(V'\right)^2
\ .
\label{EOMVn}
\ee
\par
This differential equation is obviously hard to solve analytically for a general source.
We will therefore expand the field $V$ up to first order in the coupling
$\qf$~\footnote{Since Eq.~\eqref{EOMVn} is obtained from a Lagrangian defined
up to first order in $\qf$, higher-order terms in the solution would not be meaningful.},
\be
V(r)
=
V_{(0)}(r)
+
\qf\, V_{(1)}(r)
\ ,
\label{Vexp}
\ee
and solve Eq.~\eqref{EOMVn} order by order.
In particular, we have
\be
\triangle V_{(0)}
=
4\,\pi\,\qb\,\frac{\lp}{\mpl}\,\rho
\ ,
\label{EOMVn0}
\ee
which, when $q_B=1$, is just the Poisson Eq.~\eqref{PoissonVN} for the Newtonian potential
and
\be
\triangle V_{(1)}
=
2\left(V_{(0)}'\right)^2
\ ,
\label{EOMVn1}
\ee
which gives the correction at first order in $\qf$.
\par
To linear order in $q_\Phi$, the on-shell Hamiltonian~\eqref{NewtHam} is also replaced by
\be
H[V]
&\!\!=\!\!&
-L[V]
\notag
\\
&\!\!\simeq\!\!&
4\,\pi
\int_0^\infty 
r^2 \,\d r 
\left\{
-\frac{V}{2}
\left[
\qb\,\rho+\frac{\qf\,\mpl}{2\,\pi\,\lp}\left(V'\right)^2
\right]
+\qb\, \rho\,V
- \frac{\qf\,\mpl}{2\,\pi\,\lp}\,V\left( V' \right)^2 
\right\}
\notag
\\
\!\!&\simeq&\!\!
2\,\pi \int_0^\infty \d r \, r^2
\left[
\qb\, \rho\, V \left(1-4\,\qf \, V \right)
-\qf\,\frac{3\,\mpl}{2\,\pi\,\lp}\,V \left(V'^2 \right)
\right]
\ ,
\label{HamV}
\ee
where we used Eq.~\eqref{EOMVn}.
In the following, we will still denote the on-shell contribution containing the matter density $\rho$ with
\be
U_{\rm BG}
=
2\,\pi\, \qb
\int_0^\infty 
r^2\,\d r 
\,\rho
\left[V_{(0)}
+
\qf
\left(
V_{(1)}-4\,V^2_{(0)}
\right)
\right]
+
\mathcal{O}(\qf^2)
\ ,
\label{UbgG}
\ee
which reduces to the Newtonian $U_{\rm N}$ in Eq.~\eqref{Un} for $\qb=1$ and $\qf=0$,
and the rest as
\be
U_{\rm GG}
=
-3\,\qf\,\frac{\lp}{\mpl}
\int_{0}^\infty
r^2\,\d r\,
V_{(0)} \left( V_{(0)}' \right)^2
+\mathcal{O}(\qf^2)
\ .
\label{UggG}
\ee
\subsection{Classical solutions}
We will now study the general classical solutions to Eqs.~\eqref{EOMVn0}
and \eqref{EOMVn1}.
Since we are interested in static and spherically symmetric sources,
it is convenient to consider eigenfunctions of the Laplace operator,
\be
\triangle
j_0(k\,r)
=
-k^2 \, j_0(k\,r)
\ ,
\label{Lmodes}
\ee
that is, the spherical Bessel function of the first kind
\be
j_0(k\,r)
=
\frac{\sin(k\,r)}{k\,r}
\ ,
\label{j0}
\ee
which enjoys the normalisation
\be
4\,\pi\int_0^\infty 
r^2\,\d r\, j_0(p\,r) \, j_0(k\,r)
=
\frac{2\,\pi^2}{k^2}\,\delta(p-k)
\ .
\label{Lortho}
\ee
Assuming the matter density is a smooth function of the radial coordinate,
we can project it on the above modes,
\be
\tilde\rho(k)
=
4\,\pi
\int_0^\infty
r^2\,\d r\,j_0(k\,r)\,\rho(r)
\ ,
\label{rhok}
\ee
and likewise
\be
\tilde V_{(n)}(k)
=
4\,\pi
\int_0^\infty
r^2\,\d r\,j_0(k\,r)\,V_{(n)}(r)
\ .
\label{Vnkgen}
\ee
Inverting these expressions, one obtains the expansions in Laplacian eigenfunctions,
\be
f(r)
=
\int_0^\infty
\frac{k^2\,\d k}{2\,\pi^2}\,
j_0(k\,r)\,\tilde f(k)
\ ,
\label{frgen}
\ee
in which we used  
\be
\int \frac{\d^3 k}{(2\,\pi)^3}
=
\int_0^\infty \frac{k^2\,\d k}{2\,\pi^2}
\ ,
\label{d3p}
\ee
since all our functions only depend on the radial momentum $k\ge 0$.
\par
The zero-order Eq.~\eqref{EOMVn0} in momentum space reads
\be
\tilde V_{(0)}(k)
=
-4\,\pi\,\qb\,\frac{\lp\,\tilde\rho(k)}{\mpl\,k^2}
\ ,
\label{V0kgen}
\ee
which can be inverted to yield the solution
\be
V_{(0)}(r)
=
-2\,\qb\,\frac{\lp}{\mpl}
\int_0^\infty
\frac{\d k}{\pi}\,
j_0(k\,r)\,\tilde\rho(k)
\ .
\ee
The r.h.s.~of Eq.~\eqref{EOMVn1} can then be written as
\be
2\left( V'_{(0)}(r) \right)^2
=
\qb^2 \, \frac{8\,\lp^2}{\mpl^2}
\left(
\int_0^\infty \frac{k \, \d k}{\pi} \, j_1(k\,r)\, \tilde \rho(k)
\right)^2
\ ,
\label{Jgclass}
\ee
where we used Eq.~\eqref{V0kgen} and
\be
\left[j_0(k\,r)\right]'
=
-k\,j_1(k\,r)
\ .
\ee
The first-order Eq.~\eqref{EOMVn1} is however easier to solve directly in coordinate space usually.
\par
For example, for a point-like source of mass $M_0$, whose density is given by
\be
\rho
=
M_0\,\delta^{(3)}({\bf x})
=
\frac{M_0}{4\,\pi\,r^2}\,\delta(r)
\ ,
\label{delta}
\ee
one finds 
\be
\tilde \rho(k)
=
M_0\,
\int_0^\infty
\d r\,j_0(k\,r)\,\delta(r)
=
M_0
\ ,
\ee
and Eq.~\eqref{V0kgen} yields the Newtonian potential outside a spherical source of mass
$M_0$ (for $\qb=1$), that is
\be
V_{(0)}(r)
=
-2\,\qb\,\frac{\lp\,M_0}{\mpl\,r}
\int_0^\infty
\frac{\d z}{\pi}\,
j_0(z)
=
-\qb\,\frac{\lp\,M_0}{\mpl\,r}
\ .
\label{VNdelta}
\ee
Note that this solution automatically satisfies the regularity condition
\be
\lim_{r\to\infty}
V_{(0)}(r)
=0
\ .
\label{limInf}
\ee
Next, for $r>0$, one has
\be
2\left( V'_{(0)}(r) \right)^2
\!\!&=&\!\!
\qb^2 \, \frac{8 \, \lp^2\, M_0^2}{\mpl^2\,r^4}
\left(
\int_0^\infty 
\frac{z\, \d z}{\pi} \, j_1(z)
\right)^2
\notag
\\
\!\!&=&\!\!
\qb^2 \, \frac{2 \, \lp^2\, M_0^2}{\mpl^2 \, r^4}
\ ,
\label{VPrime2}
\ee
and Eq.~\eqref{EOMVn1} admits the general solution
\be
V_{(1)}
=
A_1
-\qb\,\frac{\lp\,M_1}{\mpl\,r}
+\qb^2\frac{\lp^2\,M_0^2}{\mpl^2\,r^2}
\ .
\ee
On imposing the same boundary condition~\eqref{limInf} to $V_{(1)}$, one obtains $A_1=0$.
The arbitrary constant $M_1$ results in a (arbitrary) shift of the ADM mass,
\be
M=M_0+\qf\,M_1
\ ,
\ee
and one is therefore left with the potential
\be
V
=
-\qb\,\frac{\lp\,M}{\mpl\,r}
+\qf\,\qb^2\frac{\lp^2\,M^2}{\mpl^2\,r^2}
+\mathcal{O}(\qf^2)
\ .
\label{Vdelta}
\ee
This expression matches the expected post-Newtonian form~\eqref{Upost} at large $r$
for $\qb=\qf=1$.
It also clearly shows the limitation of the present approach:
at small $r$, the post-Newtonian correction $V_{(1)}$ grows faster than $V_{(0)}=V_{\rm N}$ and
our perturbative approach will necessarily break down.
\par
We can also evaluate the potential energy~\eqref{HamV} generated by the point-like source. 
The baryon-graviton energy~\eqref{UbgG} of course diverges, but we can regularise the matter
density~\eqref{delta} by replacing $\delta(r)\to \delta(r-r_0)$, where $0<r_0\ll \lp\, M_0/\mpl$.
We then find
\be
U_{\rm BG}
\simeq
-\qb^2\, \frac{\lp \, M_0\,M}{2\,\mpl \, r_0}
-\qb^3\, \qf\, \frac{3\,\lp^2 \, M^3}{2\,\mpl^2\, r_0^2}
\ .
\label{Ubgdelta}
\ee
With the same regularisation, we obtain the graviton-graviton energy
\be
U_{\rm GG}
\simeq
-3\,\qf\,\frac{\lp}{\mpl}
\int_{r_0}^\infty
r^2\,\d r\,
V_{(0)} \left( V_{(0)}' \right)^2
=
\qb^3\, \qf\, \frac{3\,\lp^2\, M^3}{2\,\mpl^2 \, r_0^2}
\ ,
\label{Uggdelta}
\ee
which precisely cancels against the first order correction to $U_{\rm BG}$ in Eq.~\eqref{Ubgdelta},
and
\be
U
=
U_{\rm BG}
+
U_{\rm GG}
=
-\qb^2\, \frac{\lp \, M_0\,M}{2\,\mpl \, r_0}
\ .
\label{Udelta}
\ee
Of course, for $r\simeq r_0\ll \lp\, M_0/\mpl$, the post-Newtonian
term in Eq.~\eqref{Vdelta} becomes much larger than the Newtonian contribution,
which pushes the above $U_{\rm BG}$ and $U_{\rm GG}$ beyond the regime of
validity of our approximations. 
Nonetheless, it is important to notice that, given the effective Lagrangian~\eqref{LagrV},
the total gravitational energy~\eqref{Udelta} for a point-like source will never vanish
and the maximal packing condition~\eqref{maxpack} cannot be realised.
This is consistent with the concept of corpuscular black holes as quantum objects
with a (very) large spatial extensions $R\sim\Rh$.   
\par
For the reasons above, we shall next study extended distributions of matter, which
will indeed lead to different, more sensible results within the scope of our approach.
\subsection{Homogeneous matter distribution}
\label{SShomogeneous}
For an arbitrary matter density, it is hopeless to solve the equation~\eqref{EOMVn1}
for $V_{(1)}$ analytically.
Let us then consider the very simple case in which $\rho$ is uniform inside a sphere
of radius $R$,
\be
\rho(r)
=
\frac{3\, M_0}{4\,\pi\, R^3} \, \Theta(R-r)
\ ,
\label{HomDens}
\ee
where $\Theta$ is the Heaviside step function and
\be
M_0
=
4\,\pi
\int_0^\infty
r^2\,\d r\,\rho(r)
\ .
\ee
For this matter density, we shall now solve Eqs.~\eqref{EOMVn0} and \eqref{EOMVn1}
with boundary conditions that ensure $V$ is regular both at the origin $r=0$ and infinity, 
that is
\be
V_{(n)}'(0)
=
\underset{r\to\infty}{\lim}\,V_{(n)}(r)
=
0
\label{cond0inf}
\ ,
\ee
and smooth across the border $r=R$,
\be
\underset{r\to R^-}{\lim} V_{(n)}(r)
=
\underset{r\to R^+}{\lim} V_{(n)}(r)
\ ,
\quad
\underset{r\to R^-}{\lim} V_{(n)}'(r)
=
\underset{r\to R^+}{\lim} V_{(n)}'(r)
\ .
\label{Boundary}
\ee
The solution to Eq.~\eqref{EOMVn0} inside the sphere is then given by
\be
V_{(0) \rm in}(r)
=
\qb\,\frac{\lp \, M_0}{2\,\mpl\,R^3}
\left( r^2-3\,R^2 \right)
\ee
while outside 
\be
V_{(0) \rm out}(r)
=
-\qb\,\frac{\lp \, M_0}{\mpl \,r}
\ ,
\ee
which of course equal the Newtonian potential for $\qb=1$.
\par
\begin{figure}[t]
\centering
\includegraphics[width=10cm]{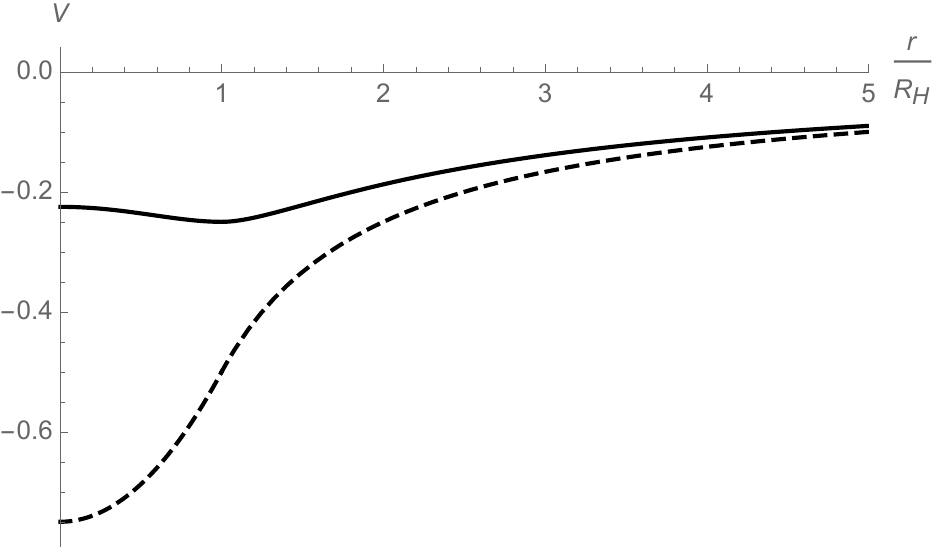}
\caption{Potential to first order in $\qf$ (solid line) vs Newtonian potential (dashed line)
for $R=2\,\lp\,M/\mpl\equiv \Rh$ and $\qb=\qf=1$.}
\label{PPN1}
\end{figure}

At first order in $\qf$ we instead have
\be
V_{(1) \rm in}(r)
=
\qb^2 \, \frac{\lp^2\,M_0^2}{10\,\mpl^2\,R^6}
\left(
r^4-15\,R^4
\right)
\label{V1inH}
\ee
and
\be
V_{(1) \rm out}(r)
=
\qb^2\, \frac{\,\lp^2\,M_0^2}{5\,\mpl^2\,R}\,
\frac{5\,R-12\,r}{r^2}
\ .
\label{V1outH}
\ee
The complete outer solution to first order in $\qf$ is thus given by
\be
V_{\rm out}(r)
=
-\qb\,\frac{\lp\,M_0}{\mpl\,r}
\left(1+\qf\,\qb\,\frac{12\,\lp\,M_0}{5\,\mpl\,R}\right)
+\qb^2\,\qf\,\frac{\lp^2\,M_0^2}{\mpl^2\,r^2}
+\mathcal{O}(\qf^2)
\ .
\ee
From this outer potential, we see that, unlike for the point-like source,
we are left with no arbitrary constant and the ADM mass is determined
as
\be
M
=
M_0
\left(1+\qf\,\qb\,\frac{12\,\lp\,M_0}{5\,\mpl\,R}\right)
+\mathcal{O}(\qf^2)
\ ,
\ee
and, replacing this expression into the solutions, we finally obtain
\be
V_{\rm in}(r)
&\!\!=\!\!&
\qb \frac{\lp\,M}{2\,\mpl\,R^3}
\left(r^2-3\,R^2\right)
+\qb^2\,\qf\,\frac{\lp^2\,M^2}{10\,\mpl^2\,R^6}
\left(r^4-12\,R^2\,r^2+21\,R^4\right)
+\mathcal{O}(\qf^2)
\ ,
\\
V_{\rm out}(r)
&\!\!=\!\!&
-\qb\,\frac{\lp\,M}{\mpl\,r}
+\qb^2\,\qf\,\frac{\lp^2\,M^2}{\mpl^2\,r^2}
+\mathcal{O}(\qf^2)
\ .
\ee
We can now see that the outer field again reproduces the first post-Newtonian result~\eqref{Upost}
of Appendix~\ref{Apost-Newton} when $\qb=\qf=1$ (see Figs.~\ref{PPN5} and \ref{PPN1} for
two examples).
\begin{figure}[t]
\centering
\includegraphics[width=10cm]{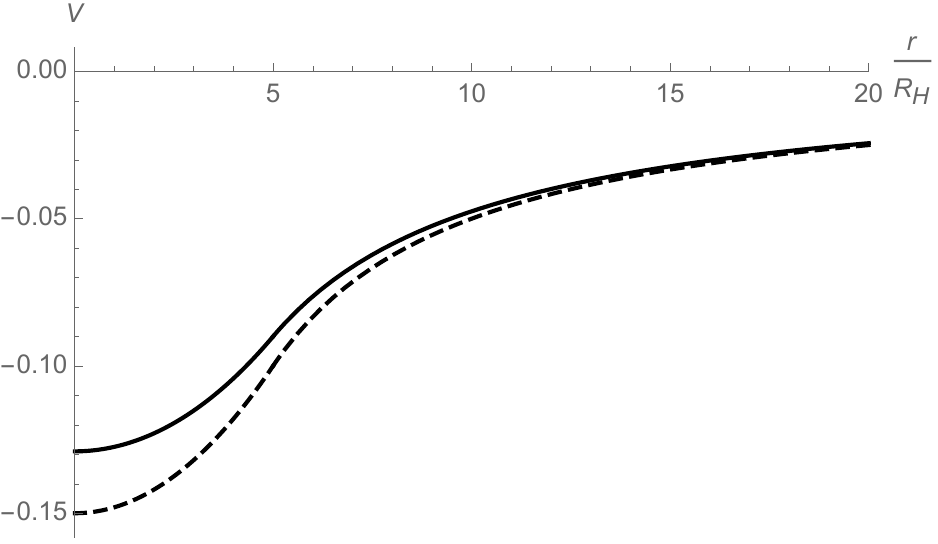}
\caption{Potential to first order in $\qf$ (solid line) vs Newtonian potential (dashed line)
for $R=10\,\lp\,M/\mpl\equiv 5\,\Rh$ and $\qb=\qf=1$.}
\label{PPN5}
\end{figure}
\par
Since the density~\eqref{HomDens} is sufficiently regular, we can evaluate the corresponding 
gravitational energy~\eqref{HamV} without the need of a regulator. 
The baryon-graviton energy~\eqref{UbgG} is found to be
\be
U_{\rm BG}(R)
&\!\!=\!\!&
2\,\pi\,\qb
\int_0^R 
r^2 \,\d r\,
\rho
\left[
V_{(0) \rm in}
+
\qf
\left(
V_{(1) \rm in}
-4\,V_{(0) \rm in}^2
\right)
\right]
+\mathcal{O}(\qf^2)
\notag
\\
&\!\!=\!\!&
-\qb^2\,\frac{3\,\lp\,M^2}{5\,\mpl\,R}
-\qb^3\,\qf\,\frac{267\,\lp^2\,M^3}{350\,\mpl^2\,R^2}
+\mathcal{O}(\qf^2)
\notag
\\
&\!\!\equiv\!\!&
U_{(0)\rm BG}(R)
+\qf\,U_{(1) \rm BG}(R)
+\mathcal{O}(\qf^2)
\ ,
\ee
where $U_{(0)\rm BG}$ is just the Newtonian contribution (for $\qb=1$) and
$U_{(1) \rm BG}$ the post-Newtonian correction.
Analogously, the self-sourcing contribution~\eqref{UggG} gives
\be
U_{\rm GG}(R)
&\!\!=\!\!&
-3\,\qf\,\frac{\mpl}{\lp}
\left[
\int_0^R
r^2\, \d r \,
V_{(0) \rm in}
\left(V'_{(0)\rm in} \right)^2 
+
\int_R^\infty
r^2\,\d r\,
V_{(0)\rm out}\left(V'_{(0)\rm out} \right)^2
\right] 
+\mathcal{O}(\qf^2)
\notag
\\
&\!\!=\!\!&
\qb^3\,\qf\,\frac{153\,\lp^2\,M_0^3}{70\,\mpl^2\,R^2}
+\mathcal{O}(\qf^2)
\ .
\ee
Since now $U_{\rm GG}>\qf\,|U_{(1) \rm BG}|$, adding the two terms together yields
the total gravitational energy
\be
U(R)
=
-\qb^2\,\frac{3\,\lp\,M^2}{5\,\mpl\,R}
+\qb^3\,\qf\,\frac{249\,\lp^2\,M^3}{175\,\mpl^2\,R^2}
+\mathcal{O}(\qf^2)
\ ,
\ee
which appears in line with what was estimated in Ref.~\cite{Casadio:2016zpl}:
the (order $\qf$) post-Newtonian energy is positive, and would equal the Newtonian contribution for
a source of radius
\be
R
\simeq
\frac{83\,\lp\,M}{35\,\mpl}
\simeq
1.2\,\Rh
\ ,
\ee
where se wet $\qb=\qf=1$.
One has therefore recovered the ``maximal packing'' condition~\eqref{maxpack}
of Refs.~\cite{DvaliGomez} in the limit $R\sim\Rh$ from a regular matter distribution.
However, note that, strictly speaking, the above value of $R$ falls outside the regime of validity 
of our approximations.
\subsection{Gaussian matter distribution}
\label{SSgaussian}
As an example of even more regular matter density, we can consider a Gaussian distribution of
width $\sigma$,
\be
\rho(r)
=
\frac{M_0\,e^{-\frac{r^2}{\sigma^2}}}{\pi^{3/2}\,\sigma^3} 
\ ,
\label{GaussDens}
\ee
where again
\be
M_0
=
4\,\pi
\int_0^\infty
r^2\,\d r\,\rho(r)
\ .
\ee
Let us remark that the above density is essentially zero for $r\gtrsim R\equiv 3\,\sigma$,
which will allow us to make contact with the previous case.
\par
\begin{figure}[t]
\centering
\includegraphics[width=10cm]{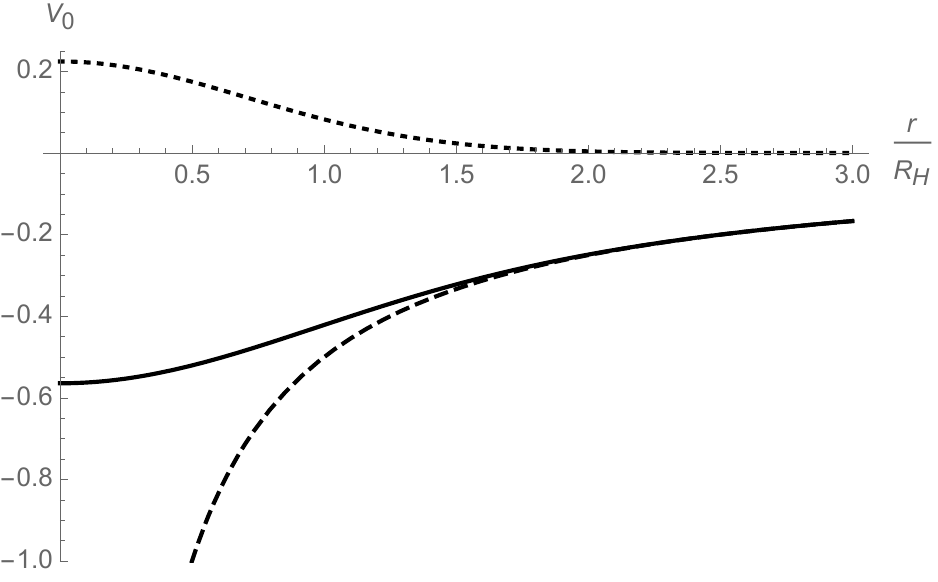}
\caption{Newtonian potential (solid line) for Gaussian matter density with $\sigma=2\,\lp\,M_0/\mpl$
(dotted line) vs Newtonian potential (dashed line) for point-like source of mass $M_0$
(with $\qb=1$).}
\label{V0G}
\end{figure}

For this matter density, we shall now solve Eqs.~\eqref{EOMVn0} and \eqref{EOMVn1}
with the boundary conditions~\eqref{cond0inf} that ensure $V$ is regular both at the
origin $r=0$ and at infinity.
We first note that Eq.~\eqref{rhok} yields
\be
\tilde\rho(k)
=
M_0\,e^{-\frac{\sigma^2\,k^2}{4}}
\ ,
\ee
from which
\be
V_{(0)}(r)
&\!\!=\!\!&
-2\,\qb\,\frac{\lp\,M_0}{\mpl}
\int_0^\infty
\frac{\d k}{\pi}\,
j_0(k\,r)\,e^{-\frac{\sigma^2\,k^2}{4}}
\notag
\\
&\!\!=\!\!&
-\qb\,\frac{\lp\,M_0}{\mpl\,r}\,{\rm Erf}(r/\sigma)
\ .
\ee
For a comparison with the analogous potential generated by a point-like source with
the same mass $M_0$, see Fig.~\ref{V0G}.
For $r\gtrsim R=3\,\sigma=3\,\Rh/2$, the two potentials are clearly indistinguishable,
whereas $V_{(0)}$ looks very similar to the case of homogeneous matter for $0\le r<R$
(see Fig.~\ref{PPN5}). 
\begin{figure}[h]
\centering
\includegraphics[width=10cm]{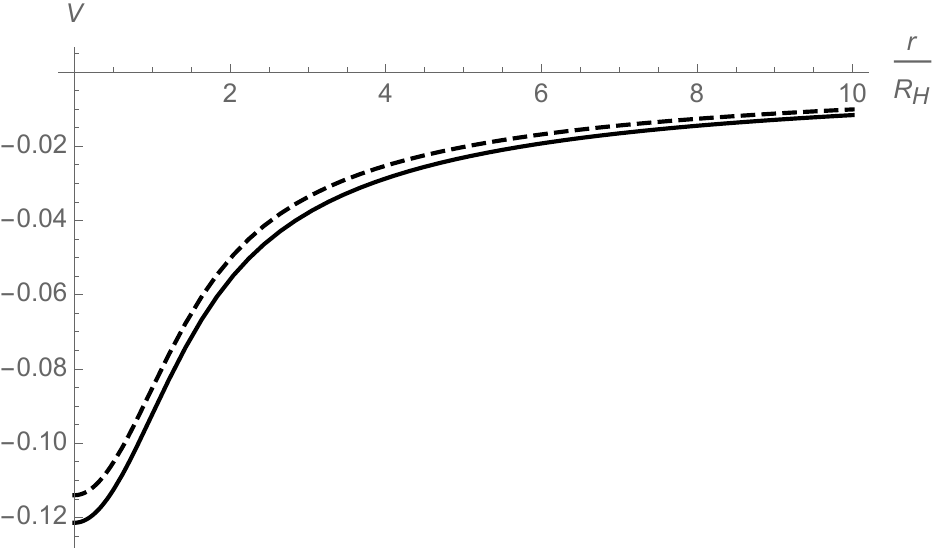}
\caption{Potential up to first order in $\qf$ (solid line) vs Newtonian potential (dashed line)
for Gaussian matter density with $\sigma=2\,\lp\,M/\mpl\equiv \Rh$ (with $\qb=\qf=1$).}
\label{V1G}
\end{figure}
\par
The first-order equation~\eqref{EOMVn1} now reads
\be
\triangle V_{(1)}
=
2\,\qb^2\,\frac{\lp\,M_0^2}{\mpl^2\,r^4}
\left[
{\rm Erf}(r/\sigma)-\frac{2\,r}{\sqrt{\pi}\,\sigma}\,e^{-\frac{r^2}{\sigma^2}}
\right]^2
\equiv
2\,\qb^2\,\frac{\lp\,M_0^2}{\mpl^2}\,
G(r)
\ ,
\label{EOMV1G}
\ee
and we note that 
\be
G(r)
\simeq
\left\{
\begin{array}{lcl}
\strut\displaystyle{\frac{16\,r^2}{9\,\pi\,\sigma^6}}
&
{\rm for}
&
r\to 0
\\
\\
\strut\displaystyle{\frac{1}{r^4}}
&
{\rm for}
&
r\to \infty
\ ,
\end{array}
\right.
\ee
which are the same asymptotic behaviours one finds for a homogenous source of size $R\sim\sigma$.
We can therefore expect the proper solution to Eq.~\eqref{EOMV1G} behaves like
Eq.~\eqref{V1inH} for $r\to 0$ and \eqref{V1outH} for $r\to \infty$.
In fact, one finds
\be
V_{(1)}
=
2\,\qb^2\, \frac{\lp^2\,M_0^2}{\mpl^2}
\left\{
\frac{\left[\erf\!\left(\frac{r}{\sigma}\right)\right]^2-1}{\sigma^2}
-
\frac{\sqrt{2}\,\erf\!\left(\sqrt{2}\,\frac{r}{\sigma}\right)}{\sqrt{\pi}\,\sigma\,r} 
+
\frac{\left[\erf\!\left(\frac{r}{\sigma}\right)\right]^2}{2\,r^2}
+
\frac{2\,e^{-\frac{r^2}{\sigma^2}}\,\erf\left(\frac{r}{\sigma}\right)}
{\sqrt{\pi}\,\sigma\,r}
\right\}
\ ,
\label{V1}
\ee
in which we see the second term in curly brackets again leads to a shift in the ADM mass,
\be
M
=
M_0
\left(
1+
\qb\,\qf\,\frac{2\,\sqrt{2}\,\lp\,M_0}{\sqrt{\pi}\,\mpl\,\sigma} 
\right)
\ ,
\label{MadmG}
\ee
while the third term reproduces the usual post-Newtonian potential~\eqref{Upost}
for $r\gg\sigma$.
For an example of the complete potential up to first order in $\qf$, see Fig.~\ref{V1G}.
Note that for the relatively small value of $\sigma$ used in that plot, the main effect of $V_{(1)}$
in Eq.~\eqref{V1} is to increase the ADM mass according to Eq.~\eqref{MadmG}, which lowers
the total potential significantly with respect to the Newtonian curve for $M=M_0$ shown in Fig.~\ref{V0G}.
\section{Quantum linear field and coherent ground state}
\setcounter{equation}{0}
\label{Squantum}
We are now going to see how one can reproduce the previous classical results in a quantum
theory.
We will proceed by canonically quantising a suitably rescaled potential field, and then
identifying the quantum state which yields expectation values close to the classical expressions.
\par
A canonically normalised scalar field $\Phi$ has dimensions of $\sqrt{{\rm mass}/{\rm length}}$,
while the potential $V$ is dimensionless.
We therefore define 
\be
\Phi
=
\sqrt{\frac{\mpl}{\lp}} \, V
\ ,
\qquad
\jb
=
4\,\pi\,\sqrt{\frac{\lp}{\mpl}} \, \rho
\ ,
\label{RescPhi}
\ee
and replace these new quantities in Eq.~\eqref{LagrV}.
After rescaling the whole Lagrangian~\eqref{LagrV} by a factor of $4\,\pi$, in order to have
a canonically normalised kinetic term, we obtain the scalar field Lagrangian 
\be
L[\Phi]
=
4\,\pi \int_0^\infty
r^2\, \d r 
\left[
\frac{1}{2}\,\Phi\,\Box \Phi
-\qb\, \jb\,\Phi
\left(1-2\,\qf\,\sqrt{\frac{\lp}{\mpl}} \,\Phi \right)
+2\, \qf\, \sqrt{\frac{\lp}{\mpl}}\,
\left(\partial_\mu \Phi \right)^2 \Phi
\right]
\ ,
\ee
where we again assumed $\Phi=\Phi(t,r)$.
\par
As usual, we define the quantum field operators starting from the ``free'' theory, corresponding
to $\qb=\qf=0$, that is we will employ normal modes of the equation
\be
\Box
\Phi
=
0
\ .
\label{box}
\ee
In particular, since we are interested in static and spherically symmetric states, we can again
employ the eigenfunctions~\eqref{Lmodes} of the Laplace operator, and define the time-dependent
modes
\be
u_k(t,r)
=
j_0(k\,r)\,e^{i\,\omega\,t}
\ ,
\label{modes}
\ee
which satisfy
\be
4\,\pi\int_0^\infty 
r^2\,\d r\, u^*_p(t,r) \, u_k(t,r)
=
\frac{2\,\pi^2}{k^2}\,\delta(p-k)
\ .
\label{ortho}
\ee
Upon replacing~\eqref{modes} into Eq.~\eqref{box}, one of course obtains
the mass-shell relation $\omega=k$, so that the field operator and its conjugate
momentum are respectively given by
\be
\hat\Phi(t,r)
=
\int_0^\infty \frac{k^2\,\d k}{2\,\pi^2}\,
\sqrt{\frac{\lp\,\mpl}{2\,k}}\,
j_0(k\,r)
\left(
\a_k \, e^{i\, k\, t} + \ac_k \, e^{-i\, k\, t}
\right)
\ ,
\ee
and
\be
\hat \Pi(t,r)
=
i\,\int_0^\infty
\frac{k^2\,\d k}{2\,\pi^2}
\sqrt{\frac{\lp\,\mpl\,k}{2}}\,
j_0(k\,r)
\left(
\a_k \, e^{i\, k\, t} - \ac_k \, e^{-i\, k\, t}
\right)
\ ,
\ee
where the creation and annihilation operators satisfy
\be
\left[ \a_h,\ac_k \right]
=
\frac{2\,\pi^2}{k^2}\,\delta(h-k)
\ ,
\label{CommLadder}
\ee
and we again used Eq.~\eqref{d3p}.
\subsection{Newtonian potential}
Let us now turn to Eq.~\eqref{EOMVn}, and look for a quantum state $\ket{g}$ of $\Phi$ which
reproduces the classical solution.
First of all, we will consider the Newtonian case, that is we set $\qf=0$ and find a solution
for Eq.~\eqref{EOMVn0}.
In terms of the new variables $\Phi$ and $\jb$, this equation reads
\be
\triangle\Phi_c(r)
=
\qb\,\jb(r)
\ ,
\label{eqPhi0}
\ee
where we emphasised that we shall only consider static currents $\jb=\jb(r)$ and correspondingly
static fields.
Upon expanding~\eqref{eqPhi0} on the modes~\eqref{j0}, one finds the classical
solution in momentum space is of course given by Eq.~\eqref{V0kgen}, which now reads
\be
\tilde\Phi_c(k)
=
-\qb\,\frac{\tilde\jb(k)}{k^2}
\ ,
\ee
with $\tilde J_{\rm B}^*(k)=\tilde\jb(k)$ from the reality of $\jb(r)$, and analogously
$\tilde\Phi_c^*(k)=\tilde\Phi_c(k)$.
We then define the coherent state 
\be
\a_k
\ket{ g}
=
e^{i\,\gamma_k(t)}\,
g_k
\ket{g}
\ ,
\ee
where
\be
g_k
=
\sqrt{\frac{k}{2\,\lp\mpl}}\,\tilde\Phi_c(k)
=
-\qb\,\frac{\tilde\jb(k)}{\sqrt{2\,\lp\mpl\,k^3}}
\ .
\label{gk}
\ee
Acting on such a state with the operator $\hat\Phi$ yields the expectation value
\be
\bra{g}
\hat \Phi(t,r)
\ket{g}
&\!\!=\!\!&
\bra{g}
\int_0^\infty \frac{k^2\,\d k}{2\,\pi^2}\,
\sqrt{\frac{\lp\,\mpl}{2\,k}}\,
j_0(k\,r)
\left(
\a_k \, e^{i\, k\, t} + \ac_k \, e^{-i\, k\, t}
\right)
\ket{g}
\nonumber
\\
&\!\!=\!\!&
\pro{g}{g}\,
\int_0^\infty \frac{k^2\,\d k}{2\,\pi^2}\,
\sqrt{\frac{\lp\,\mpl}{2\,k}}\,
j_0(k\,r)
\left[
g_k\, e^{i\, k\, t+i\,\gamma_k(t)} + g^*_k \, e^{-i\, k\, t-i\,\gamma_k(t)}
\right]
\nonumber
\\
&\!\!=\!\!&
-\qb\,\pro{g}{g}\,
\int_0^\infty \frac{k^2\,\d k}{2\,\pi^2}\,
\frac{j_0(k\,r)}{2\,k^2}\,\tilde\jb(k)
\left[
 e^{i\, k\, t+i\,\gamma_k(t)}
+
e^{-i\, k\, t-i\,\gamma_k(t)}
\right]
\ .
\quad
\ee
Now, assuming $\pro{g}{g}=1$ and $\gamma_k(t)=-k\,t$, we finally obtain
\be
\bra{g}
\hat \Phi(t,r)
\ket{g}
&\!\!=\!\!&
-\qb\,
\int_0^\infty \frac{k^2\,\d k}{2\,\pi^2}\,
j_0(k\,r)\,\frac{\tilde\jb(k)}{k^2}
\nonumber
\\
&\!\!=\!\!&
\int_0^\infty \frac{k^2\,\d k}{2\,\pi^2}\,
j_0(k\,r)\,\tilde\Phi_c(k)
=
\Phi_c(r)
\ ,
\ee
which is exactly the classical solution to Eq.~\eqref{eqPhi0}.
\par
It is particularly important to study the normalisation of $\ket{g}$.
One can explicitly write this state in terms of the true vacuum $\ket{0}$ as  
\be
\ket{g}
=
e^{-\frac{N_{\rm G}}{2}} \exp \left\{
\int_0^\infty
\frac{k^2\,\d k}{2\,\pi^2}\,g_k\,\ac_k
\right\}
\,\ket{0}
\ ,
\label{cohg}
\ee
where $N_{\rm G}$ is just a normalisation factor for now.
By making use of the commutation relation~\eqref{CommLadder} and the 
well-known Baker-Campbell-Hausdorff formulas, one then obtains
\be
\pro{g}{g}
\!\!&=&\!\!
e^{-N_{\rm G}}
\bra{0}
\exp \left\{
\int_0^\infty
\frac{p^2\,\d p}{2\,\pi^2}
\,g_p^*\,\a_p
\right\}
\exp \left\{
\int_0^\infty
\frac{k^2\,\d k}{2\,\pi^2}
\,g_k\,\ac_k
\right\}
\ket{0}
\notag
\\
\!\!&=&\!\!
e^{-N_{\rm G}}
\bra{0}
\exp \left\{
\int_0^\infty
\frac{p^2\,\d p}{2\,\pi^2}
\int_0^\infty
\frac{k^2\,\d k}{2\,\pi^2}
\,g_p\,g_k\,\left[ \a_p,\ac_k \right]
\right\}
\ket{0}
\notag
\\
\!\!&=&\!\!
e^{-N_{\rm G}}
\exp \left\{
\int_0^\infty
\frac{k^2\,\d k}{2\,\pi^2}
\,g^2_k\right\}
\ ,
\ee
so that
\be
N_{\rm G}
=
\int_0^\infty
\frac{k^2\,\d k}{2\,\pi^2}
\, g^2_k
=
\bra{g}
\int_0^\infty
\frac{k^2\,\d k}{2\,\pi^2}
\,\ac_k\,\a_k\,
\ket{g}
\ ,
\ee
and $N_{\rm G}$ is shown to precisely equal the total occupation number of modes in the state $\ket{g}$.
This quantity typically diverges, as we can show with a simple example.
\par
Let us consider the point-like source~\eqref{delta}, for which we have
\be
g_k
=
-\qb\,\frac{4\,\pi\,M_0}{\mpl\,\sqrt{2\,k^3}}
\ .
\ee
The general treatment above shows that the zero-order field will exactly equal the (suitably rescaled)
Newtonian potential~\eqref{VNdelta}, and
\be
N_{\rm G}
=
\qb^2\,\frac{4\,M_0^2}{\mpl^2}
\int_{k_0}^\Lambda
\frac{\d k}{k}
=
\qb^2\,\ln\!\left(\frac{\Lambda}{k_0}\right)\frac{4\,M_0^2}{\mpl^2}
\ ,
\label{Ndelta}
\ee
where we introduced both a infrared cut-off $k_0$ and a ultraviolet cut-off $\Lambda$ to regularise
the divergences.
The latter originates from the source being point-like, which allows for modes of infinitely large
momentum, and is usually not present when one considers regular matter densities.
The former is instead due to assuming the source lives in an infinite volume or, equivalently,
is eternal so that its static gravitational field extends to infinite distances.
\par
Had we considered a source of mass $M$ with finite size $R$, we can anticipate that one would
typically find 
\be
N_{\rm G}
\sim
\frac{M^2}{\mpl^2}\,\ln\!\left(\frac{R_\infty}{R}\right)
\ ,
\label{NgR}
\ee
where $R_\infty=k_0^{-1} \gg R$ denotes the size of the universe within which the gravitational
field is static.
It is of paramount importance to note that $N_{\rm G}$ depends on $R$ much less than it does
on the mass $M$, since
\be
\frac{\d N_{\rm G}}{N_{\rm G}}
\sim
2\,\frac{\d M}{M}
-\frac{1}{\ln(R_\infty/R)}\,\frac{\d R}{R}
\ ,
\label{dndelta}
\ee
and the effect of the variation in the source size $R$ can be made arbitrarily small by
simply choosing a very large $R_\infty$. 
This results can in fact be confirmed explicitly by employing the Gaussian source~\eqref{GaussDens},
that is
\be
\tilde \jb(k)
=
4\,\pi\, M_0\, \sqrt{\frac{\lp}{\mpl}} 
\, e^{-\frac{\sigma^2\, k^2}{4}}
\ ,
\ee
from which 
\be
N_{\rm G}
=
\qb^2 \, \frac{4\,M_0^2}{\mpl^2}
\int_{k_0}^\infty
\frac{\d k}{k}\,
e^{-\frac{\sigma^2\, k^2}{4}}
=
\qb^2 \, \frac{2\,M_0^2}{\mpl^2}\,
\Gamma\!\left(0, \frac{\sigma^2}{R_\infty^2}\right)
\ ,
\ee
where we again introduced a cut-off $k_0= 1/(2\,R_\infty)$ and 
\be
\Gamma(0,x)
=
\int_x^\infty \frac{\d t}{t} \, e^{-t}
\ee
is the (lower) incomplete gamma function.
The relative variation,
\be
\frac{\d N_{\rm G}}{N_{\rm G}}
=
2\,\frac{\d M_0}{M_0}
-
\frac{2\,e^{-\sigma^2/R_\infty^2}}{\Gamma\left(0, \sigma^2/R_\infty^2\right)}
\,\frac{\d \sigma}{\sigma}
\ ,
\ee
shows once more that the number of quanta in the coherent state is much more influenced
by changes in the bare mass of the source than it is by changes in the width $\sigma$,
for the arbitrary cut-off $R_\infty$ may be taken much larger than $\sigma$.
Moreover, since
\be
\Gamma\!\left(0, \frac{\sigma^2}{R_\infty^2}\right)
\simeq
2\,\ln\!\left(\frac{R_\infty}{\sigma}\right)
\ ,
\ee
we see that the estimate in Eq.~\eqref{NgR} is actually confirmed by taking $R\simeq \sigma$. 
\subsection{Post-Newtonian correction}
Having established that 
\be
\sqrt{\frac{\lp}{\mpl}}
\bra{g}
\hat \Phi(t,r)
\ket{g}
=
V_{\rm N}(r)
=
V_{(0)}(r)
\ ,
\ee
solves Eq.~\eqref{EOMVn0}, we can tackle Eq.~\eqref{EOMVn1}, which we now rewrite as
\be
\triangle V_{(1)}
=
2\,\frac{\lp}{\mpl}\,
\bra{g}\left(\hat\Phi'\right)^2\ket{g}
\ .
\label{EOMVn1q}
\ee
In the above,
\be
\hat\Phi'(t,r)
=
-\int_0^\infty \frac{k^2\,\d k}{2\,\pi^2}\,
\sqrt{\frac{\lp\,\mpl}{2\,k}}\,k\,
j_1(k\,r)
\left(
\a_k \, e^{i\, k\, t} + \ac_k \, e^{-i\, k\, t}
\right)
\ ,
\ee
so that
\be
2\,\frac{\lp}{\mpl}
\bra{g}
\left(\hat \Phi'\right)^2
\ket{g}
&\!\!=\!\!&
2\,\lp^2
\int_0^\infty
\frac{p^{5/2}\,\d p}{2\,\sqrt{2}\,\pi^2}\,
\int_0^\infty
\frac{k^{5/2}\,\d k}{2\,\sqrt{2}\,\pi^2}\,
j_1(p\,r)\,j_1(k\,r)
\nonumber
\\
&&
\qquad
\times
\bra{g}
\left(
\a_p \, e^{i\, p\, t} + \ac_p \, e^{-i\, p\, t}
\right)
\left(
\a_k \, e^{i\, k\, t} + \ac_k \, e^{-i\, k\, t}
\right)
\ket{g}
\nonumber
\\
&\!\!=\!\!&
2\,\lp^2
\int_0^\infty
\frac{p^{5/2}\,\d p}{2\,\sqrt{2}\,\pi^2}\,
\int_0^\infty
\frac{k^{5/2}\,\d k}{2\,\sqrt{2}\,\pi^2}\,
j_1(p\,r)\,j_1(k\,r)
\left(
4\,g_p\,g_k+\left[\a_p,\ac_k\right] e^{i\,(p-k)\,t}
\right)
\nonumber
\\
&\!\!=\!\!&
8\,\lp^2
\left[
\int_0^\infty
\frac{k^{5/2}\,\d k}{2\,\sqrt{2}\,\pi^2}
\,j_1(k\,r)\,g_k
\right]^2
+
2\,\lp^2
\int_0^\infty
\frac{k^3\,\d k}{4\,\pi^2}
\left[j_1(k\,r)\right]^2
\nonumber
\\
&\!\!\equiv\!\!&
J_g+J_0
\ .
\label{expecPhi2}
\ee
Note that the (diverging) term denoted by $J_0$ is a purely vacuum contribution
independent of the quantum state and we can simply discard it by imposing the
normal ordering in the expectation value above.
From the expression~\eqref{gk} of the eigenvalues $g_k$, with the rescaling~\eqref{RescPhi}
for the matter density, one can immediately see that $J_g$ equals the classical 
expression~\eqref{Jgclass}, that is
\be
2\,\frac{\lp}{\mpl}
\bra{g}
\left(\hat \Phi'\right)^2
\ket{g}
=
2\,\left(V_{(0)}'\right)^2
\ ,
\label{JgV0}
\ee
for any matter distribution.
This shows that the coherent state $\ket{g}$ obtained from the Newtonian potential is
indeed a very good starting point for our perturbative quantum analysis.
\par
We should now determine a modified coherent state $\ket{g'}$, such that 
\be
\sqrt{\frac{\lp}{\mpl}}\,
\bra{g'}\hat \Phi\ket{g'}
\simeq
V_{(0)}
+
\qf\,V_{(1)}
\ ,
\label{expP}
\ee
where all expressions will be given to first order in $\qf$ from now on.
Like we expanded the classical potential in Eq.~\eqref{Vexp}, we can also write
\be
\ket{g'}
\simeq
\mathcal{N}
\left(
\ket{g}
+
\qf
\ket{\delta g}
\right)
\ ,
\ee
with
\be
\a_k\ket{g'}
\simeq
g_k\ket{g}+\qf\,\delta g_k\ket{\delta g}
\ ,
\ee
and the normalisation constant
\be
|\mathcal{N}|^2
\simeq
1 - 2\, \qf \, {\rm Re}\,\pro{\delta g}{g}
\ .
\ee
Upon replacing these expressions, we obtain
\be
\bra{g'}\hat \Phi\ket{g'}
&\!\!\simeq\!\!& 
\left(1 - 2\, \qf \, {\rm Re}\,\pro{\delta g}{g}\right)
\left(
\bra{g}\hat \Phi\ket{g} + 
2\, \qf \,{\rm Re}\,\bra{\delta g} \hat \Phi \ket{g}
\right)
\nonumber
\\
&\!\!\simeq\!\!&
\bra{g}\hat \Phi\ket{g} +
2\, \qf \, {\rm Re}\,\bra{\delta g} \hat \Phi \ket{g}
-
2\, \qf \, \bra{g}\hat \Phi\ket{g}\,{\rm Re}\, \pro{\delta g}{g}
\ ,
\ee
and Eq~\eqref{expP} yields
\be 
{\rm Re}\bra{\delta g} \hat \Phi \ket{g}
-
\bra{g}\hat \Phi\ket{g}\,{\rm Re}\pro{\delta g}{g} 
=
\sqrt{\frac{\mpl}{\lp}}\,
\frac{V_{(1)}}{2} 
\ .
\ee
By applying the Laplacian operator on both sides, we finally get
\be
\frac{\triangle \left({\rm Re}\bra{\delta g} \hat \Phi \ket{g} \right)}
{{\rm Re}\pro{\delta g}{g}}
=
\triangle \bra{g}\hat \Phi\ket{g}
+
\sqrt{\frac{\lp}{\mpl}} \,
\frac{\bra{g}\left(\hat \Phi'\right)^2\ket{g}}
{{\rm Re}\pro{\delta g}{g}}
\ ,
\label{eq-schifo}
\ee 
where we used Eqs.~\eqref{EOMVn1q}.
\par
The above equation relates each eigenvalue $\delta g_k$ to all of the $g_p$'s in the
Newtonian coherent state, which obviously makes solving it very complicated.
We will instead estimate the solution by following the argument of Ref.~\cite{Casadio:2016zpl} 
that was summarised in the Introduction.
Namely, we assume most of the $N_{\rm G}$ gravitons are in one mode of wavelength
$\lambda_{\rm G}\simeq R$~\cite{DvaliGomez}, so that
\be 
\hat \Phi
\simeq
\sqrt{\lp\, \mpl}\,\bar k^{3/2}\,\Delta\bar k\,j_0 (\bar{k}\,r)\,
\left(\a _{\bar{k}}+\ac _{\bar{k}}\right)
\ ,  
\ee
where $\bar k\simeq R^{-1}\simeq\Delta\bar k$, and we neglect numerical factors of order one.
In particular, we have
\be 
\bra{g}\hat \Phi\ket{g} 
&\!\!\simeq\!\!&
\sqrt{\lp\, \mpl}\,\bar k^{3/2}\,\Delta\bar k\,j_0 (\bar{k}\,r)\,
g_{\bar k}
\\
\bra{\delta g}\hat \Phi\ket{g}
&\!\!\simeq\!\!&
\pro{\delta g}{g}\,
\sqrt{\lp\, \mpl}\,\bar k^{3/2}\,\Delta\bar k\,j_0 (\bar{k}\,r)
\left(g_{\bar k} + \delta g_{\bar k}\right)
\ ,
\ee
and
\be
\bra{g}
\left(\hat \Phi'\right)^2
\ket{g} 
\simeq
\lp\,\mpl\,\bar k^5\,(\Delta\bar k)^2\,
j_1^2 (\bar{k}\,r)\,g_{\bar{k}}^2
\ ,
\ee
where we again subtracted the vacuum term $J_0$ from Eq.~\eqref{expecPhi2}.
Plugging these results into Eq.\eqref{eq-schifo} finally yields
\be 
\delta g _{\bar k}
\simeq
-\lp\,\bar k^{3/2}\,\Delta\bar k\,g_{\bar k}^2
\sim
-\lp\,\bar k^{5/2}\,g_{\bar k}^2
\ .
\ee
\par
For instance, for the point-like source~\eqref{delta}, one obtains
\be
\delta g _{\bar k}
\sim
-\frac{\lp\,M_0^2}{\mpl^2\,\bar k^{1/2}}
\sim
\frac{\lp\,M_0}{\mpl\,r_0}\,g_{\bar k}
\sim
\frac{\Rh}{r_0}\,g_{\bar k}
\ ,
\ee
where we set the characteristic size of the source $R\sim r_0$, the latter being the same
ultra-violet cut-off we introduced for computing the (diverging) classical gravitational
energy~\eqref{Udelta}.  
For $r_0\ll \Rh$, this result clearly falls outside the range of our approximations, since 
$\delta g_{\bar k}\gg g_{\bar k}$ (of course, we assume $\qb\sim\qf\sim 1$).
For the Gaussian source~\eqref{GaussDens}, we instead obtain
\be
\delta g _{\bar k}
\sim
\frac{\lp\,M_0^2}{\mpl^2\,\bar k^{1/2}}\,e^{-\frac{\sigma^2\,\bar k^2}{2}}
\sim
\frac{\Rh}{\sigma}\,g_{\bar k}
\ ,
\label{dgG}
\ee
having set $\bar k\simeq R^{-1}\sim \sigma^{-1}$.
We then see the perturbation $\delta g _{\bar k}\ll g _{\bar k}$ when the source
is much more extended than its gravitational radius, which is indeed consistent with the
classical results we are trying to reproduce quantum mechanically.
\section{Conclusions and outlook}
\setcounter{equation}{0}
\label{Sconc}
Starting from the Einstein-Hilbert action in the weak field and non-relativistic approximations,
we have built an effective quantum description of the static gravitational potential up to first
post-Newtonian order.
The detailed calculations presented here substantially support the energy balance outlined in
Ref.~\cite{Casadio:2016zpl}, and the consequent derivation of the maximal packing
condition~\eqref{maxpack}, which is a crucial ingredient for corpuscular models of
black holes~\cite{DvaliGomez}.
Moreover, our analysis can help to clarify a number of subtle aspects we would like to finally comment on.
\par
Let us begin from the expression~\eqref{NgR} for the number $N_{\rm G}$ of (virtual) gravitons in the
Newtonian potential.
Although it remains true $N_{\rm G}$ mainly depends on the mass of the static source,
we found it also (weakly) depends on the ratio $R/R_\infty$ between the size of the source 
and the size of the region within which the gravitational potential is static.  
Such a dependence becomes negligible for an ideal static system (with $R_\infty\to\infty$),
but could play a much bigger role in a dynamical situation when the source evolves in time
and the extension of the outer region of static potential is comparable to $R$.
In fact, the number $N_{\rm G}$ in Eq.~\eqref{NgR} vanishes for $R_\infty\simeq R$ and grows 
logarithmically with $R_\infty$, meaning that the (Newtonian) coherent state $\ket{g}$ becomes
(logarithmically) more and more populated as the region of static potential extends further and further
away from the source.
\par
It would also be tempting to consider the case $R=\Rh$ and relate the second term in Eq.~\eqref{dndelta}
to logarithmic corrections for the Bekenstein-Hawking entropy of black holes.
We have however noted repeatedly that a source of size $R\lesssim \Rh$ usually falls outside the regime of
validity of our approximations.
Nonetheless, from the classical point of view, nothing particularly wrong seems to happen
in the limiting case $R\simeq\Rh$, except the very equality~\eqref{maxpack} that gives support to
the corpuscular model of black holes now occurs precisely in this borderline condition.
That $R\sim \Rh$ becomes critical for our description is further made clear by the estimate~\eqref{dgG}
of quantum corrections $\ket{\delta g}$ to the coherent state $\ket{g}$ that reproduces the Newtonian
potential, since the corrections must become comparable to the Newtonian part for $\sigma\sim R\to\Rh$.
Whether this is in full agreement with the post-Newtonian description of General Relativity or it instead 
signals a breakdown of the classical picture near the threshold of black hole formation will require
a much more careful analysis. 
We leave this seemingly very relevant topic of quantum perturbations, along with the role of matter
pressure (which we totally neglected here), for future works.
\section*{Acknowledgments}
We are indebted to W.~M\"uck for very useful comments.
R.C.~and A.~Giusti~are partially supported by the INFN grant FLAG.
The work of R.C., A.~Giugno and A.~Giusti has also been carried out in the framework
of activities of the National Group of Mathematical Physics (GNFM, INdAM).
\appendix
\section{Post-Newtonian potential}
\setcounter{equation}{0}
\label{Apost-Newton}
In order to derive the post-Newtonian correction to the usual Newtonian 
potential from General Relativity, we consider a test particle of mass $m$ freely
falling along a radial direction in the Schwarzschild space-time around a
source of mass $M$. 
\par
The Schwarzschild metric in standard form is given by~\footnote{In this Appendix, we will
use units with $\gn=1$ for simplicity.}
\be
\d s^2
=
-
\left(1-\frac{2\,M}{\tilde r}\right)\d \tilde t^2
+
\left(1-\frac{2\,M}{\tilde r}\right)^{-1}\d \tilde r^2
+
\tilde r^2\,\d\Omega^2
\ ,
\ee
and the radial geodesic equation for a massive particle turns out to be
\be
\frac{\d^2 \tilde r}{\d \tau^2}
=
-\frac{M}{\tilde r^2}
\ ,
\label{eq}
\ee
which looks formally equal to the Newtonian expression, but where $\tilde r$ is the areal radial coordinate
related to the Newtonian radial distance $r$ by
\be
\d r
=
\frac{\d \tilde r}{\sqrt{1-\frac{2\,M}{\tilde r}}}
\ .
\ee
Moreover, the proper time $\tau$ of the freely falling particle is related to the Schwarzschild
time $\tilde t$ by 
\be
\d \tau
=
\left(1-\frac{2\,M}{\tilde r}\right)
\frac{m}{E}\,\d \tilde t
\ ,
\label{taut}
\ee
where $E$ is the conserved energy of the particle.
We thus have
\be
\frac{\d^2 \tilde r}{\d \tilde t^2}
&\!\!=\!\!&
-
\frac{M}{\tilde r^2}
\left(1-\frac{2\,M}{\tilde r}\right)^{2}
\left[
\frac{m^2}{E^2}
-2\left(1-\frac{2\,M}{\tilde r}\right)^{-3}
\left(\frac{\d \tilde r}{\d \tilde t}\right)^2
\right]
\ .
\ee
\par
Next, we expand the above expressions for $M/r\simeq M/\tilde r\ll 1$ (weak field) and $|\d\tilde r/\d\tilde t|\ll 1$
(non-relativistic regime).
In order to keep track of small quantities, it is useful to introduce a parameter $\epsilon>0$
and replace
\be
\frac{M}{\tilde r}
\to
\epsilon\,
\frac{M}{\tilde r}
\ ,
\qquad
\frac{\d\tilde r}{\d \tilde t}
\to
\epsilon\,
\frac{\d \tilde r}{\d \tilde t}
\ .
\ee
From the non-relativistic limit, it also follows that $E=m+\mathcal{O}(\epsilon^2)$
and any four-velocity
\be
u^\mu
=
\left(
1+\mathcal{O}(\epsilon^2),
\epsilon\,\frac{\d\vec x}{\d\tilde t}
+
\mathcal{O}(\epsilon^2)
\right)
\ ,
\ee
so that the acceleration is also of order $\epsilon$,
\be
\frac{\d^2 x^\mu}{\d \tau^2}
=
\epsilon
\left(
0,\frac{\d^2\vec x}{\d \tilde t^2}
\right)
+
\mathcal{O}(\epsilon^2)
\ .
\ee
We then have
\be
\epsilon\,
\frac{\d^2 \tilde r}{\d \tilde t^2}
=
-
\epsilon\,
\frac{M}{\tilde r^2}
\left(1-\epsilon\,\frac{2\,M}{\tilde r}\right)^{2}
\left[
1+\mathcal{O}(\epsilon^2)
-2\,\left(1-\epsilon\,\frac{2\,M}{\tilde r}\right)^{-3}
\epsilon^2\left(\frac{\d \tilde r}{\d \tilde t}\right)^2
\right]
\ ,
\label{eqx}
\ee
and
\be
r
\simeq
\int
\left(1+\epsilon\,\frac{M}{\tilde r}+\epsilon^2\,\frac{3\,M}{2\,\tilde r^2}\right)
\d \tilde r
\simeq
\tilde r
\left[
1
-
\epsilon\,\frac{M}{\tilde r}\,\log\!\left(\epsilon\,\frac{M}{\tilde r}\right)
-\epsilon^2\,\frac{3\,M^2}{2\,\tilde r^2}
+
\mathcal{O}(\epsilon^3)
\right]
\ .
\ee
Since
\be
r
=
\tilde r
+\mathcal{O}\left(\epsilon \log\epsilon\right)
\ ,
\ee
it is clear that Eq.~\eqref{eqx} to first order in $\epsilon$ reproduces the Newtonian dynamics,
\be
\frac{\d^2 r}{\d \tilde t^2}
\simeq
\frac{\d^2 \tilde r}{\d \tilde t^2}
\simeq
-
\frac{M}{r^2}
\ .
\ee
The interesting correction comes from including the next order.
In fact, we have
\be
\epsilon\,
\frac{\d^2\tilde r}{\d \tilde t^2}
=
-
\epsilon\,
\frac{M}{r^2}
+\epsilon^2\,\frac{4\,M^2}{r^3}
+\mathcal{O}\left(\epsilon^2 \log\epsilon\right)
\ ,
\ee
or, neglecting terms of order $\epsilon^2 \log\epsilon$ and higher, and then setting
$\epsilon=1$,
\be
\frac{\d^2r}{\d \tilde t^2}
=
-\frac{M}{r^2}
+\frac{4\,M^2}{r^3}
=
-\frac{\d}{\d r}
\left(
-\frac{M}{r}
+\frac{2\,M^2}{r^2}
\right)
\ .
\ee
The correction to the Newtonian potential would therefore appear to be
\be
V
=
\frac{2\,M^2}{r^2}
\ ,
\label{Upost2}
\ee
but one step is stil missing.
\par
Instead of the Schwarzschild time $\tilde t$, let us employ the proper time $t$ of static observers
placed along the trajectory of the falling particle, that is
\be
\d t
=
\left(1-\frac{2\,M}{r}\right)^{1/2}
\d \tilde t
\label{tstat}
\ .
\ee
From Eq.~\eqref{taut} we obtain 
\be
\frac{\d}{\d \tau}
=
\left(1-\frac{2\,M}{r}\right)^{-1/2}
\frac{E}{m}\,\frac{\d}{\d t}
\ ,
\ee
and
Eq.~\eqref{eq} then becomes
\be
\frac{\d^2\tilde r}{\d t^2}
&\!\!=\!\!&
-
\frac{M}{r^2}
\left(1-\frac{2\,M}{\tilde r}\right)
\left[
\frac{m^2}{E^2}
-\left(1-\frac{2\,M}{\tilde r}\right)^{-2}
\left(\frac{\d\tilde r}{\d t}\right)^2
\right]
\ .
\ee
Introducing like before the small parameter $\epsilon$ yields
\be
\epsilon\,
\frac{\d^2\tilde r}{\d t^2}
=
-
\epsilon\,
\frac{M}{\tilde r^2}
\left(1-\epsilon\,\frac{2\,M}{\tilde r}\right)
\left[
1+\mathcal{O}(\epsilon^2)
-\left(1-\epsilon\,\frac{2\,M}{\tilde r}\right)^{-2}
\epsilon^2\left(\frac{\d\tilde r}{\d t}\right)^2
\right]
\ ,
\label{eqxx}
\ee
The first order in $\epsilon$ is of course the same.
However, up to second order, one obtains
\be
\epsilon\,
\frac{\d^2 r}{\d t^2}
&\!\!=\!\!&
-
\epsilon\,\frac{M}{r^2}
+\epsilon^2\,\frac{2\,M^2}{r^3}
+\mathcal{O}\left(\epsilon^2 \log\epsilon\right)
\ ,
\ee
which yields the correction to the Newtonian potential
\be
V
=
\frac{M^2}{r^2}
\ .
\label{Upost}
\ee
This is precisely the expression following from the isotropic form of the Schwarzschild
metric~\cite{weinberg}, and the one we will consider as our reference term throughout this
paper.
\section{Linearised Einstein-Hilbert action at NLO}
\setcounter{equation}{0}
\label{EHNLO}
We shall here consider the Einstein-Hilbert and the matter actions in the non-relativistic limit,
up to NLO in the weak field expansion
\be
g_{\mu\nu}=\eta_{\mu\nu}+\epsilon\, h_{\mu\nu}
\ .
\ee 
Unlike the main text, the parameter $\epsilon$ is here shown explicitly in order to keep
track of the different orders in the expansions
\be
X
=
\sum_n
\epsilon^n\,
X_{(n)}
\ .
\ee
First of all, one has
\be
g^{\mu\nu}
=
\eta^{\mu\nu}
-\epsilon\, h^{\mu\nu}
+\epsilon^2 h^{\mu \lambda}\,h^\nu_\lambda
+\mathcal{O}(\epsilon^3)
\ ,
\ee
the integration measure reads
\be
\sqrt{-g}
=
1+\frac{\epsilon}{2}\,h
+\frac{\epsilon^2}{8}\left(h^2-2\,h_\mu^{\ \nu}\, h_\nu^{\ \mu}\right)
+\mathcal{O}(\epsilon^3)
\ ,
\label{detgeps}
\ee
and the scalar $\mathcal{R}=g^{\mu\nu}\,R_{\mu\nu}$ is obtained 
from the Ricci tensor
\be
R_{\mu\nu}
=
\partial_\lambda \Gamma^\lambda_{\mu\nu}-\partial_\nu \Gamma^\lambda_{\mu\lambda}
+\Gamma^\lambda_{\lambda\rho}\Gamma^\rho_{\mu\nu}
-\Gamma^\lambda_{\nu\rho}\Gamma^\rho_{\mu\lambda}
\ ,
\ee
provided one has computed the Christoffel symbols
\be
\Gamma^\lambda_{\mu\nu}
\simeq
\frac{\epsilon}{2}
\left(
\eta^{\lambda\rho}-\epsilon\, h^{\lambda\rho}+\epsilon^2\, h^{\lambda\sigma}\,h^\rho_\sigma
\right)
\left(
\partial_\mu h_{\rho\nu}+\partial_\nu h_{\rho\mu}-\partial_\rho h_{\mu\nu}
\right)
\ .
\ee
\par
In the de~Donder gauge~\eqref{ddgg}, the effective Lagrangian~\eqref{LagrNewt} for the classical
Newtonian field appears as the sum of two terms,
\be
L[V_{\rm N}]
=
\epsilon^2\,L_{\rm FP}
+
\epsilon\,L_{\rm M}
\ ,
\ee
with the gravitational part given by the massless Fierz-Pauli action~\cite{deser}
\be
L_{\rm FP}
\!\!&=&\!\!
\frac{\mpl}{16\, \pi\,\lp}
\int 
\d^3 x
\left(
\frac{1}{2}\,\partial_\mu h\,\partial^\mu h
-\frac{1}{2}\,\partial_\mu h_{\nu\sigma}\,\partial^\mu h^{\nu\sigma}
+\partial_\mu h_{\nu\sigma}\,\partial^\nu h^{\mu\sigma}
-\partial_\mu h\,\partial_\sigma h^{\mu\sigma}
\right)
\notag
\\
\!\!&=&\!\!
\frac{\mpl}{16\, \pi\,\lp}
\int 
\d^3 x
\left(
\partial_\mu h_{\nu\sigma}\,\partial^\nu h^{\mu\sigma}
-\frac{1}{2}\,\partial_\mu h_{\nu\sigma}\,\partial^\mu h^{\nu\sigma}
\right)
\notag
\\
\!\!&\simeq&\!\!
-\frac{\mpl}{32\, \pi\,\lp}
\int 
\d^3 x
\,\partial_\mu h_{00}\,\partial^\mu h_{00}
\notag
\\
\!\!&=&\!\!
-4\,\pi
\int_0^\infty 
r^2\,\d r\,
\frac{\mpl}{8\, \pi\,\lp}\left(V'\right)^2
\ ,
\label{LFP}
\ee
where we used the de~Donder gauge~\eqref{ddgg} and $h_{00}=-2\,V$.
The matter Lagrangian is obtained from the matter Lagrangian density~\eqref{Lmp0}, that is
\be
L_{\rm M}
\!\!&=&\!\!
\int
\d^3 x 
\left(\sqrt{-g}\, \mathcal{L}_{\rm M}
\right)_{(1)}
\notag
\\
\!\!&\simeq&\!\!
4\,\pi
\int_0^\infty 
r^2\,\d r\,
\frac{h_{00}}{2}\,\rho
\notag
\\
\!\!&=&\!\!
-4\,\pi
\int_0^\infty 
r^2\,\d r\,
V\,\rho
\ .
\label{LM1}
\ee
Putting the two pieces together yields Eq.~\eqref{LagrNewt}.
\par
The above expressions at the Newtonian level show that the factor $\mpl/(8\,\pi\,\lp)$ must be viewed as of
order $\epsilon^{-1}$, since the Einstein tensor at order $\epsilon^{n+1}$ couples to 
the stress-energy tensor at order $\epsilon^n$.
In order to go to the next order, we must then compute third-order terms for the gravitational part
and second order terms for the matter part.
After some tedious algebra, one finds
\be
-\left(\sqrt{-g}\,\mathcal{R}\right)_{(3)}
\!\!& \simeq &\!\!
h^\mu_\nu 
\left(\partial_\mu h^\lambda_\rho\, \partial^\nu h^\rho_\lambda 
-\partial^\lambda h^\nu_\mu\, \partial_\lambda h\right)
+2\,h^\mu_\nu \,\partial_\lambda h^\rho_\mu 
\left(\partial^\lambda h^\nu_\rho- \partial^\nu h^\lambda_\rho \right)
\notag
\\
&&
-\frac{1}{2}\,h\, \partial^\mu h^\lambda_\nu\, \partial_\mu h^\nu_\lambda
+\frac{1}{4}\,h\,\partial_\mu h\, \partial^\mu h
\notag
\\
\!\!&\simeq &\!\!
-h_{00}
\left(\partial_r h_{00}\right)^2
\notag
\\
\!\!& \simeq &\!\!
V \left(V' \right)^2
\ ,
\label{Jh00}
\ee
which we notice is proportional to $-J_V$ in Eq.~\eqref{JV}, and
\be
\left(\sqrt{-g}\,\mathcal{L}_{\rm M}\right)_{(2)}
=
\frac{1}{8}\, h_{00}^2\,T_{00}
=
\frac{1}{2}\, V^2\, \rho
\ .
\label{rhoV2}
\ee
Adding all the contributions, and explicitly rescaling $\mpl/(8\,\pi\,\lp)$ by a factor of $\epsilon^{-1}$,
one obtains the action
\be
S[V]
=
4\,\pi
\int
\epsilon\,\d t
\int_0^\infty 
r^2\,\d r
\left\{
\frac{\mpl}{8\,\pi\,\lp}\, V\,\triangle V
-\rho\, V 
+\frac{\epsilon}{2}\,
\left[
\frac{\mpl}{4\,\pi\,\lp}
\left(V'\right)^2
+V\,\rho
\right]
V
\right\}
\ .
\label{LagrEps}
\ee
A few remarks are now in order.
First of all, we have derived Eq.~\eqref{LagrEps} in the de~Donder gauge~\eqref{ddgg}, which 
explicitly reads
\be
\partial_t h_{00}=0
\label{ddg}
\ee
for static configurations $h_{00}=h_{00}(r)$, and is therefore automatically satisfied in our case.
This means that the above action can be used for describing the gravitational potential
$V=V(r)$ measured by any static observer placed at constant radial coordinate $r$ (provided
test particles move at non-relativistic speed). 
In fact, there remains the ambiguity in the definition of the observer time $t$, which
in turn determines the value of $\epsilon$ in Eq.~\eqref{LagrEps}, as can be seen
by the simple fact that the time measure is $\epsilon\,\d t$. 
On the other hand, changing $\epsilon$, and therefore the time (albeit in such a way
that motions remain non-relativistic) does not affect the dynamics of the Newtonian part of the
potential, whereas the post-Newtonian part inside the curly brackets acquires a different weight. 
This is completely consistent with the expansion of the Schwarzschild metric
described in Appendix~\ref{Apost-Newton}, in which we showed that the Newtonian
potential is uniquely defined by choosing a static observer, whereas the form of the
first post-Newtonian correction varies with the specific choice of time.
\par
At this point, it is convenient to introduce the (dimensionless) matter coupling $\qb$,
originating from the stress-energy tensor, by formally rescaling $\rho \to \qb\,\rho$
in the above expressions.
Likewise, the ``self-coupling'' $\qf$ will designate terms of higher order in $\epsilon$.
In particular, we set $\epsilon= 4\,\qf$ so that the post-Newtonian potential~\eqref{Upost}
is recovered for $\qf=1$~\footnote{The post-Newtonian correction~\eqref{Upost2} can instead
be obtained for $\qf=2$.}.
With these definitions, the above action yields the Lagrangian~\eqref{LagrV}.

\begin{thebibliography}{99}
%
%
\bibitem{weinberg}
S.~Weinberg,
``Gravitation and Cosmology: Principles and Applications of the General Theory of Relativity,''
(Wiley and Sons, New York, 1972)
  %
\bibitem{faraoni}
V.~Faraoni,
Symmetry {\bf 7} (2015) 2038
[arXiv:1510.03789 [gr-qc]].
%
\bibitem{duff}
M.J.~Duff,
Phys.\ Rev.\ D {\bf 7} (1973) 2317.
%
\bibitem{donoghue}
J.F.~Donoghue, M.M.~Ivanov and A.~Shkerin,
``EPFL Lectures on General Relativity as a Quantum Field Theory,''
arXiv:1702.00319 [hep-th].
%
\bibitem{Casadio:2016zpl}
R.~Casadio, A.~Giugno and A.~Giusti,
Phys.\ Lett.\ B {\bf 763} (2016) 337.
%
\bibitem{DvaliGomez} 
G.~Dvali and C.~Gomez,
JCAP {\bf 01} (2014) 023;
``Black Hole's Information Group'',
arXiv:1307.7630;
Eur.\ Phys.\ J.\ C {\bf 74} (2014) 2752;
Phys.\ Lett.\ B {\bf 719} (2013) 419;
Phys.\ Lett.\ B {\bf 716} (2012) 240;
Fortsch.\ Phys.\  {\bf 61} (2013) 742;
G.~Dvali, C.~Gomez and S.~Mukhanov,
``Black Hole Masses are Quantized,''
arXiv:1106.5894 [hep-ph].
%
\bibitem{flassig}
D.~Flassig, A.~Pritzel and N.~Wintergerst,
Phys.\ Rev.\ D {\bf 87} (2013) 084007.
%
\bibitem{becBH}
R.~Casadio, A.~Giugno, O.~Micu and A.~Orlandi,
Phys.\ Rev.\ D {\bf 90} (2014) 084040;
R.~Casadio, A.~Giugno and A.~Orlandi,
Phys.\ Rev.\ D {\bf 91} (2015) 124069
[arXiv:1504.05356 [gr-qc]];
R.~Casadio, A.~Giugno, O.~Micu and A.~Orlandi,
Entropy {\bf 17} (2015) 6893;
F.~K\"uhnel,
Phys.\ Rev.\ D {\bf 90} (2014) 084024
[arXiv:1312.2977 [gr-qc]];
F.~K\"uhnel and B.~Sundborg,
JHEP {\bf 1412} (2014) 016
[arXiv:1406.4147 [hep-th]];
F.~K\"uhnel and B.~Sundborg,
Phys.\ Rev.\ D {\bf 90} (2014)  064025
[arXiv:1405.2083 [hep-th]].
%
\bibitem{qhbh}
R.~Casadio and A.~Orlandi,
JHEP {\bf 1308} (2013) 025
[arXiv:1302.7138 [hep-th]].
%
\bibitem{mueck}
W.~M\"uck and G.~Pozzo,
JHEP {\bf 1405} (2014) 128
[arXiv:1403.1422 [hep-th]].
%
\bibitem{kerr}
R.~Casadio, A.~Giugno, A.~Giusti and O.~Micu,
``Horizon Quantum Mechanics of Rotating Black Holes,''
arXiv:1701.05778 [gr-qc].
%
\bibitem{ruffini}
R.~Ruffini and S.~Bonazzola,
Phys.\ Rev.\  {\bf 187} (1969) 1767;
%
M.~Colpi, S.L.~Shapiro and I.~Wasserman,
 Phys.\ Rev.\ Lett.\  {\bf 57} (1986) 2485;
%
M.~Membrado, J.~Abad, A.~F.~Pacheco and J.~Sanudo,
Phys.\ Rev.\ D {\bf 40} (1989) 2736;
%
%
T.M.~Nieuwenhuizen,
Europhys.\ Lett.\  {\bf 83} (2008) 10008;
%
%
P.-H.~Chavanis and T.~Harko,
Phys.\ Rev.\ D {\bf 86} (2012) 064011.
%
\bibitem{dvali}
G.~Dvali and A.~Gu{\ss}mann,
Nucl.\ Phys.\ B {\bf 913}  (2016) 1001
[arXiv:1605.00543 [hep-th]];
G.~Dvali and A.~Gu{\ss}mann,
``Aharonov-Bohm protection of black hole's baryon/skyrmion hair,''
arXiv:1611.09370 [hep-th].
%
\bibitem{kuhnel}
F.~K\"uhnel and M.~Sandstad,
Phys.\ Rev.\ D {\bf 92} (2015) 124028
 [arXiv:1506.08823 [gr-qc]].
%
\bibitem{adm}
R.L.~Arnowitt, S.~Deser and C.W.~Misner,
Phys.\ Rev.\  {\bf 116} (1959) 1322.
%
\bibitem{bekenstein}
J.D.~Bekenstein,
Phys.\ Rev.\ D {\bf 7} (1973) 2333.
%
\bibitem{madsen88}
M.S.~Madsen,
Class.\ Quant.\ Grav.\  {\bf 5} (1988) 627;
J.~D.~Brown,
Class.\ Quant.\ Grav.\  {\bf 10} (1993) 1579
[gr-qc/9304026].
%
\bibitem{harko}
T.~Harko,
Phys.\ Rev.\ D {\bf 81} (2010) 044021
[arXiv:1001.5349 [gr-qc]].
%
\bibitem{deser}
S.~Deser,
Gen.\ Rel.\ Grav.\  {\bf 1} (1970) 9
[gr-qc/0411023];
S.~Deser,
Gen.\ Rel.\ Grav.\  {\bf 42} (2010) 641
[arXiv:0910.2975 [gr-qc]].
%
\end{thebibliography}
\end{document}